\documentclass{article}
\usepackage[utf8]{inputenc}

\usepackage{natbib}
\usepackage{graphicx}
\usepackage{sidecap}
\usepackage{authblk}

\title{Requirements for gravity measurements on the anticipated Artemis III mission}

\author[1]{Peter James}
\author[2]{Anton Ermakov}
\author[3]{Michael Sori}
\affil[1]{Baylor University}
\affil[2]{University of California, Berkeley}
\affil[3]{Purdue University}

\date{August 2020}

\begin{document}

\maketitle
\tableofcontents
\newpage

\section{Introduction}

The purpose of this document is to demonstrate the reasoning behind the specific measurement requirements in the white paper by \textit{James et al.} titled ``\textit{The value of surface-based gravity and gravity gradient measurements at the Moon's south pole with Artemis III}''.

\textbf{NOTE}: As described below, measurement requirements in practice will depend on a number of factors, including the geographic location, the shape of the local terrain, the precision to which elevation is known, and the nature of drift in the gravimeter.

\section{Measurement requirements for various geological and geophysical processes}

\subsection{Science objective \#1 -- Detect buried volcanic and tectonic structures}

A tectonic or volcanic process that produces a change in the distribution of mass underground will generally produce a gravity anomaly.  An excess of mass produces a positive ``Bouguer gravity anomaly'' (defined to be the change in gravity acceleration after taking into account elevation and the mass of topography), whereas a deficit of mass produces a negative Bouguer gravity anomaly.  The amplitude of the gravity anomaly is dependent on the size, geometry, and depth of the mass anomaly.  The gravity gradients (i.e., the second derivatives of the gravitational potential) give the spatial rate of change of the vector components of gravity acceleration.  Consequently, gravity gradients are particularly sensitive to shallow, sharp density contrasts.

\begin{figure}[h!]
\centering
\hspace*{-2cm}\includegraphics[width=1.25\textwidth]{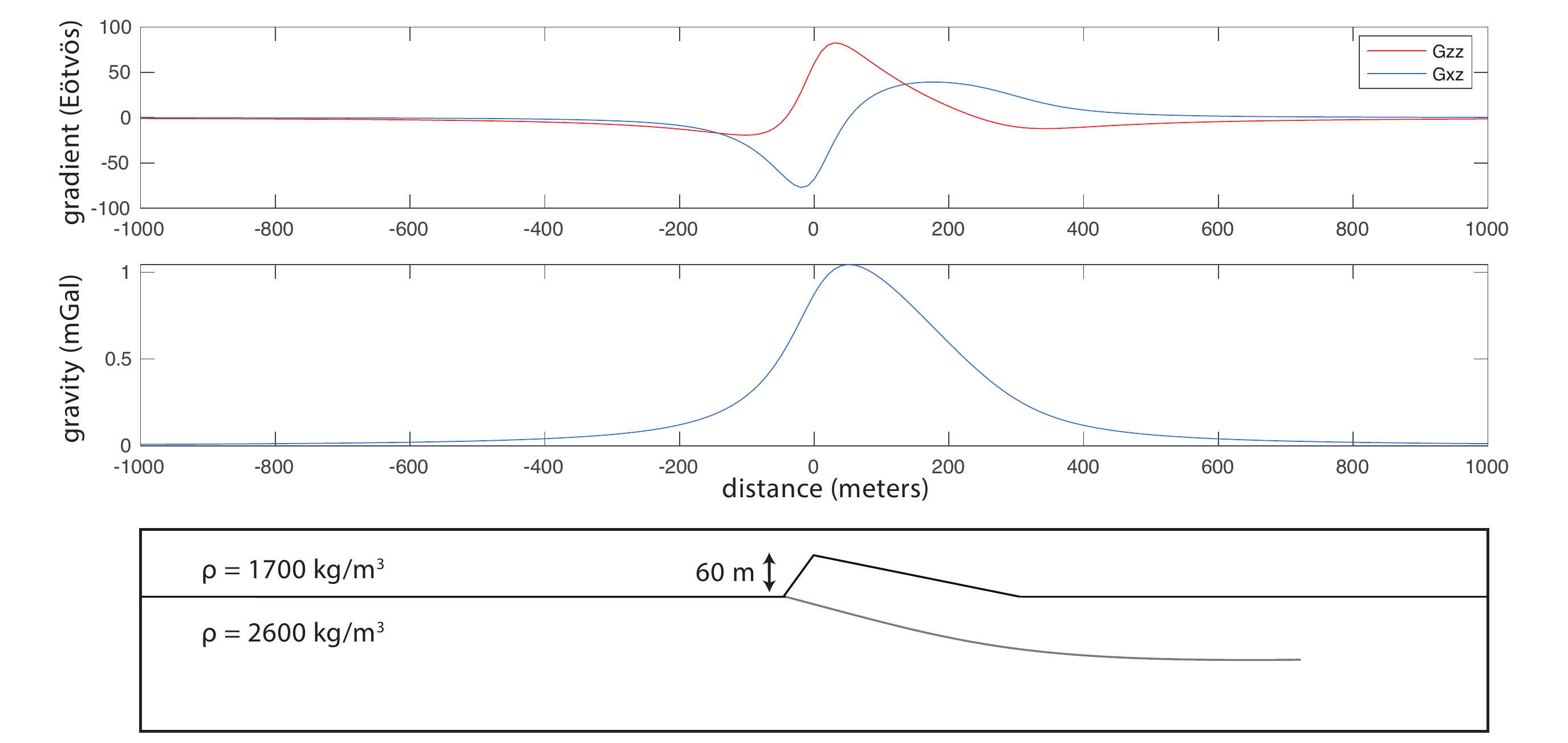}
\caption{The gravity gradients and gravity anomaly that would be generated by a tectonic scarp buried under the lunar regolith.}
\label{fig:Scarp}
\end{figure}

\begin{figure}[h!]
\centering
\hspace*{-2cm}\includegraphics[width=1.25\textwidth]{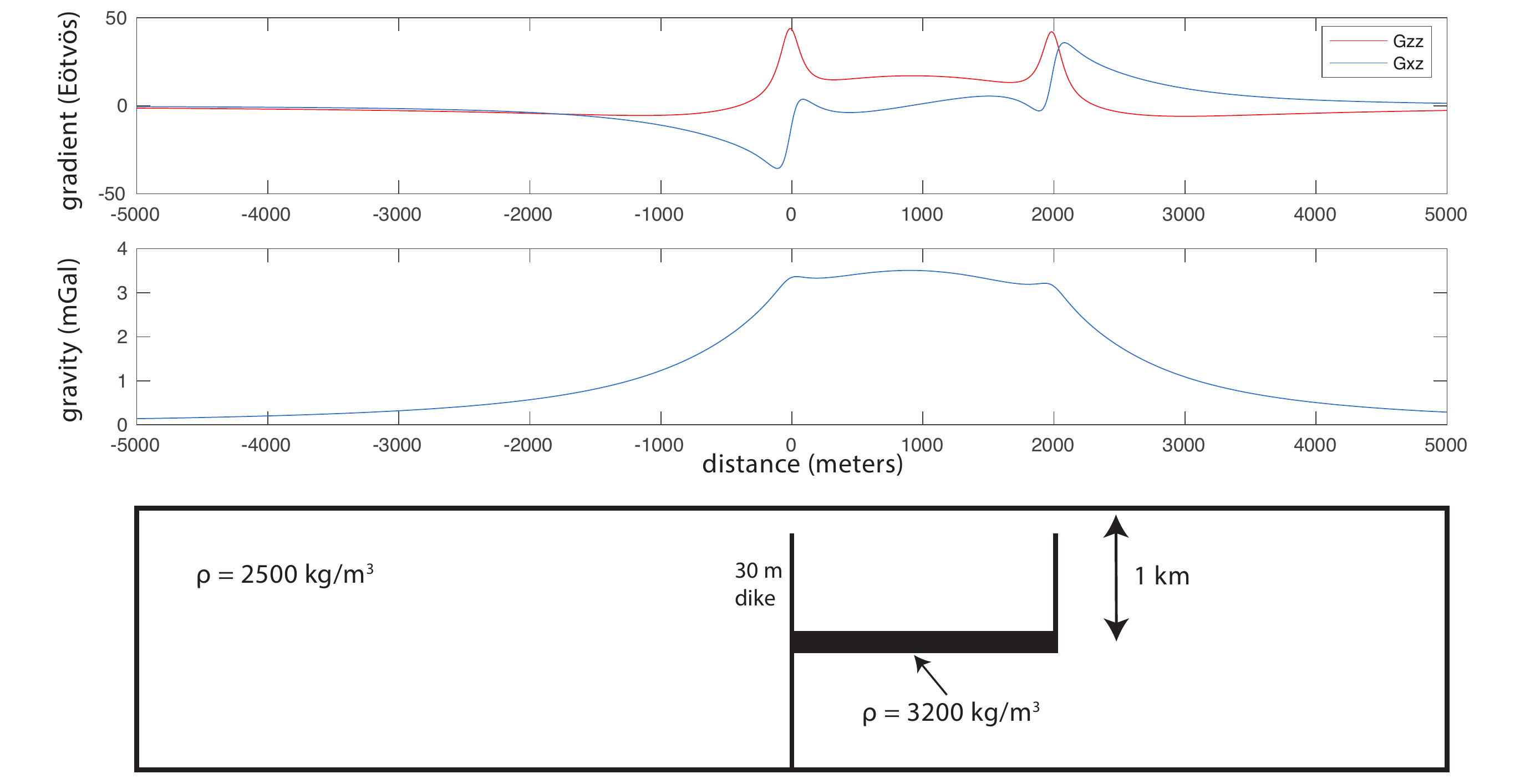}
\caption{The gravity gradients and gravity anomaly that would be generated by a pair of 30-meter-wide intrusive volcanic dikes connected by a 200-meter-tall sill.}
\label{fig:Dikes}
\end{figure}

\begin{figure}[h!]
\centering
\hspace*{-2cm}\includegraphics[width=1.25\textwidth]{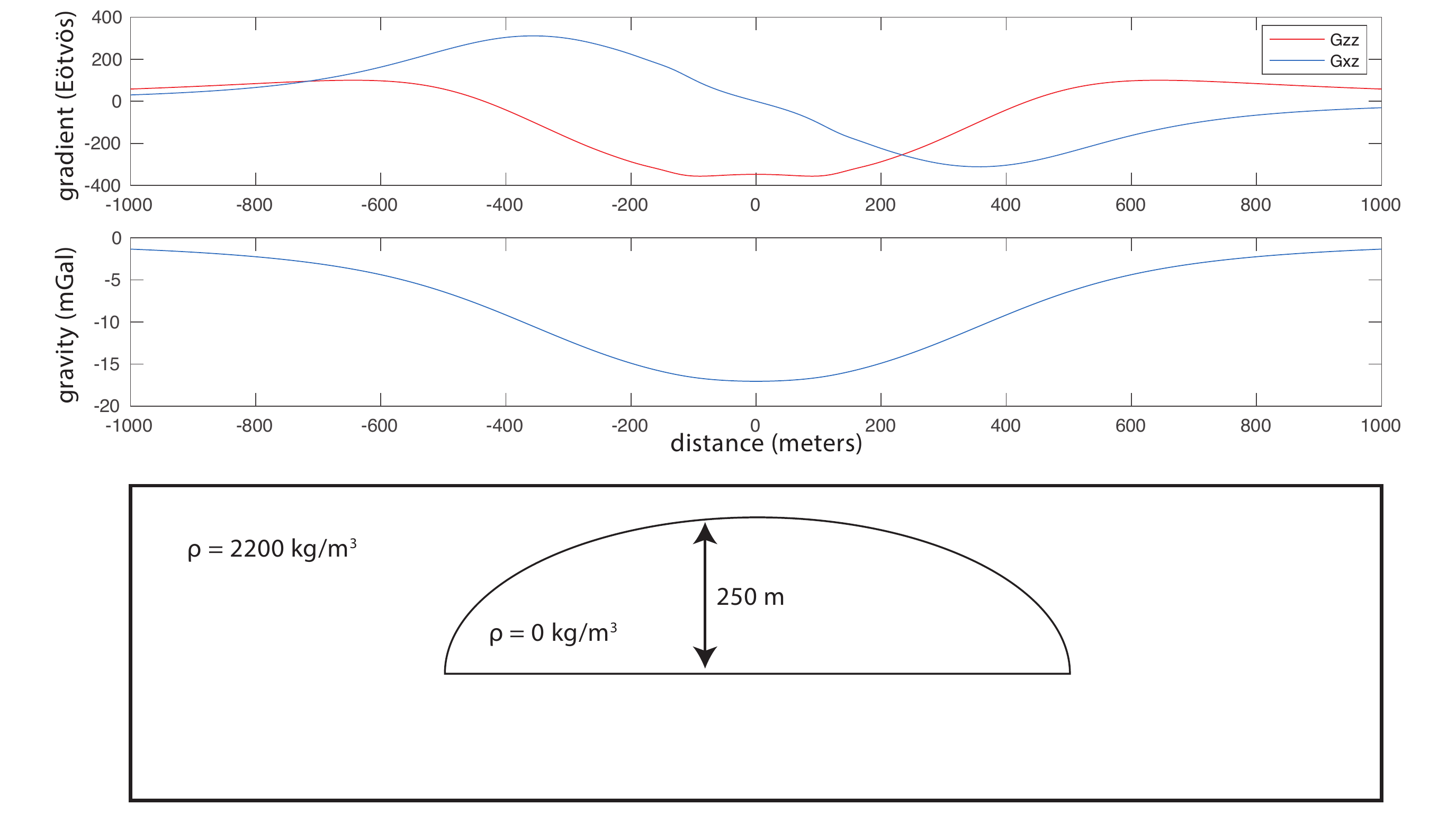}
\caption{The gravity gradients and gravity anomaly that would be generated by a lava tube that is roughly 10\% of the maximum theoretical width or 1\% of the maximum theoretical cross-sectional area \citep{blair2017structural}.}
\label{fig:LavaTube}
\end{figure}

Three such processes are illustrated in figures \ref{fig:Scarp}, \ref{fig:Dikes}, and \ref{fig:LavaTube}.  Figure \ref{fig:Scarp} illustrates a thrust fault scarp buried under the low-density regolith.  For a scarp buried at 100 meters depth with 60 meters of relief, a peak gravity anomaly of roughly 1 mGal is expected.  Meanwhile, the $G_{zz}$ and $G_{xz}$ components of the gravity gradient tensor produces maximum anomalies of roughly 80 E\"otv\"os respectively, where an E\"otv\"os is defined to be $10^{-9} $s$^{-2}$.

Figure \ref{fig:Dikes} illustrates a series of intrusive volcanic structures. Two 30-meter-wide dikes (i.e., vertical planes filled with solidified magma) are included, along with one 200-meter-thick sill (i.e., a horizontal plane filled with solidified magma) at one kilometer depth.  For intrusive magmatic bodies with these geometries and a density contrast of 700 kg / m$^{3}$ with respect to the surrounding crust, the maximum gravity anomaly is a little more than 3 mGals. The corresponding gravity gradients are maximized at roughly 40 E\"otv\"os respectively.

Finally, figure \ref{fig:LavaTube} illustrates a lava tube with a width of one kilometer and a height of 250 meters. Since lava tubes are void spaces, they represent a significant deficit of mass and produce a large negative Bouguer gravity anomaly. The maximum-amplitude gravity anomaly is roughly -18 mGals. The corresponding gravity gradients are maximized at roughly 300 E\"otv\"os respectively.

While lava tubes are suspected to exist in various locations in the lunar maria, lava tubes are not known to exist in the vicinity of the expected landing sites for Artemis III.  Consequently, we will take our measurement requirements from the other two test problems illustrated here.  The minimum gravimeter resolution that would plausibly identify near-surface features of geologic interest would be \textbf{3 mGals} (cf. Figure \ref{fig:Dikes}).  The minimum gravity gradiometer resolution that would plausibly identify near-surface features of geologic interest would be \textbf{100 E\"otv\"os} (cf. Figure \ref{fig:Scarp}).

\subsection{Science objective \#2 -- Measure porosity in the near subsurface}

When gravity is measured on short baselines with undulating topography, the resulting data may be used to estimate the bulk density of the lunar regolith.  While the bulk density of the lunar regolith is somewhat dependent on the composition of the regolith, the bulk density is primarily determined by the porosity of the regolith:

\begin{equation}
    \rho_{bulk} = (1-\phi) \rho_{grain}
\end{equation}

\noindent where $\phi$ is porosity, $\rho_{bulk}$ is the bulk density, and $\rho_{grain}$ is the average grain density of the regolith particles.

\begin{SCfigure}
\includegraphics[width=0.5\textwidth]{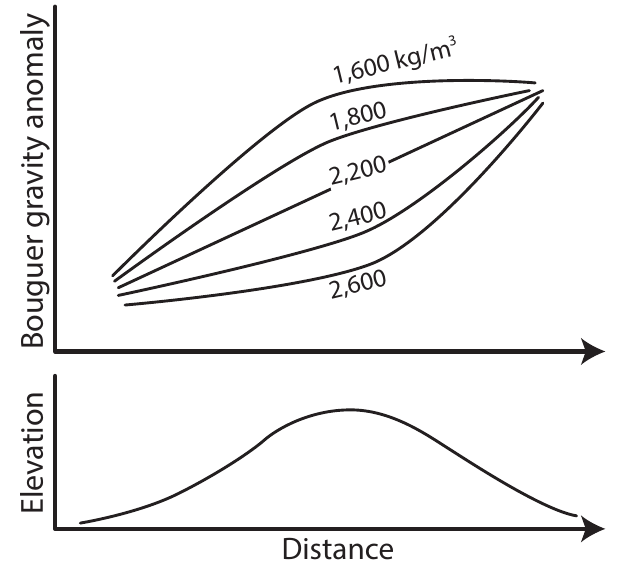}
\caption{An illustration of Nettleton's method for determining the bulk density of a terrain (modified from \citet{hinze2013gravity}). For this example, the density of the terrain has been determined to be 2200 kg/m$^3$ since that is the density that de-correlates the Bouguer gravity anomaly (top) from the topography (bottom).}
\label{fig:Nettleton}
\end{SCfigure}

In practice, the density of a terrain may be measured using the method of \citet{nettleton1939determination}. In this technique, gravity is measured at a variety of elevations in a rolling terrain.  Short-wavelength topography is generally independent of geology in the deeper crust, in which case the Bouguer anomaly would not be correlated with topography.  Therefore, when the proper bulk density for the near subsurface is chosen, the resulting Bouguer gravity anomaly should become uncorrelated with topography.  This process is illustrated in Figure \ref{fig:Nettleton}.

Bulk density varies significantly with depth in the lunar interior \citep{besserer2014grail}, and the effective depth-sensitivity of a Nettleton-type analysis is dependent on the wavelength of the topography.  When gravity from harmonic topography is upward continued by a distance $z$, that signal is attenuated by a factor of $\exp(-2\pi z /\lambda)$ where $\lambda$ is the wavelength. Therefore the characteristic sensitivity depth $z_s$ is equal to $\lambda / 2\pi$.  For example, gravity collected over topography with a wavelength of 100 meters would be sensitive to the uppermost 16 meters of the subsurface.

Measurement objectives for this scientific investigation are governed by the precision with which we would like to measure the bulk density of the near subsurface.  We will arbitrarily aim to estimate bulk density with the same precision as \citet{lewis2019surface} -- namely, 180 kg/m$^3$.  The largest source of variation in gravitational acceleration corresponds to the distance from the Moon's center of mass, so the first step in interpreting a set of raw gravity readings is to calculate the ``free-air gravity anomaly'', which is the residual gravity acceleration after accounting for the elevation of the measurement.  The free-air gravity difference $\Delta g$ corresponding to an elevation difference $\Delta h$ is, to first order, predicted by a Bouguer slab (i.e., an infinite horizontal slab) between these two elevations.  In practice, this allows us to calculate bulk density:

\begin{equation}
    \rho = \frac{\Delta g}{2\pi G\Delta h}
\end{equation}

\noindent This may be further broken down by noting that the raw gravity reading is converted to Bouguer gravity through a topographic correction. The Moon's free air gradient is 0.18 mGal/m, which must be added to a raw gravity reading:

\begin{equation}
    \rho = \frac{\Delta g_{raw} 
    +0.18 \Delta h}{2\pi G\Delta h}
\end{equation}

\noindent or

\begin{equation}
    \rho = \frac{\Delta g_{raw}}{2\pi G\Delta h} + \frac{0.18 }{2\pi G}
\end{equation}

\noindent The error in the density measurement is then determined by the error in the raw gravity reading, $E(\Delta g_{raw})$, as well as the error in the elevations, $E(\Delta h)$:

\begin{equation}
    E(\rho) \approx \frac{E(\Delta g_{raw})}{2\pi G\Delta h} + \frac{\Delta g_{raw} }{2\pi G \Delta h} \cdot \frac{E(\Delta h)}{\Delta h}
\end{equation}

From the first term, we can see that if a survey traverses 20 meters of elevation, \textbf{0.1 mGal} of error in the gravity readings would translate to 159 kg/m$^3$ of error in the calculated density.  The second term shows that the error in the density estimate is roughly proportional to the fractional error in elevation.  In other words, 1 meter of error in a survey traversing 20 meters of elevation would produce a 5\% error in a gravity reading, which would be roughly 100 kg/m$^3$ for the lunar regolith. Consequently, we recommend elevation determination better than \textbf{1 meter} (and preferably less than that).

\subsection{Science objective \#3 -- Detect water ice in the near subsurface}

     Water exists in the lunar polar regions in the form of ice \citep[see review in][]{lawrence2017tale}, but its abundance, lateral and vertical distribution, and purity is unclear.  In contrast to the relatively pure, thick, surface ice in Mercury's polar regions \citep[e.g.,][]{chabot2014images,deutsch2018constraining}, most or all of the ice present in the Moon's polar regions is likely to be patchy, mixed with rocks or regolith, and/or buried.  The reasons for the differences between polar ice on the Moon and Mercury are not fully understood.  Furthermore, such lunar ice is considered critical from an exploration standpoint, an idea that has led NASA's Human Exploration and Operations Mission Directorate to consider a more complete understanding of it to be a ``strategic knowledge gap''.  Here, we consider whether surface gravimetry can meaningfully constrain lunar ice (and therefore, lunar water).  
     
     If ice is pore-filling, it produces a positive gravity anomaly.  The lunar regolith may have substantial porosity, possibly as high as $\sim$80\% near the surface \citep{hapke2016porosity} and decreasing with depth.  Some authors speculate that ice could extend to as deep as $\sim$100 m (Cannon and Deutsch, this issue).  If pore-filling ice composes 30\% of the lunar soil by volume and extends over 100 m depth, its gravity anomaly can be estimated by the Bouguer slab approximation (justified for a surface measurement): $2\pi G \cdot 30\% \cdot 920$ kg/m$^3 \cdot$ 100 m = 1.2 mGal.  If ice takes the place of silicate material rather than pore space, it would produce a negative gravity anomaly of similar magnitude.  Therefore, gravimetry measurements with precision better than 1 mGal have the ability to meaningfully constrain ice content with respect to currently hypothesized volumes and thicknesses, whereas precision better than 0.1 mGal would constrain ice content in the more likely scenario that ice is only meters to tens of meters thick.  
     
     Gravity data with precision $<$0.1 mGal complemented by in situ grain density and other geophysical data sensitive to ice content (i.e., measurements of the subsurface electric conductivity) would constitute a strong evidence for near surface ice content variations. Mapping of grain densities would be needed to understand whether lunar ice represents a mass excess (taking the place of porosity) or mass deficit (taking the place of silicates) at all depths.  In addition, the gravity data should be correlated against such environmental factors such as the illumination conditions as surface ice should be preferentially located in permanently shadowed regions (PSRs), which are abundant within 6$^{\circ}$ of the south pole (the possible landing region of Artemis III).  It is also possible that subsurface ice is preferentially located in regions that are very cold on average, but are not PSRs and experience high annual temperature swings that pump water molecules into the subsurface \citep{schorghofer2014lunar}.  Therefore, surface gravimetry holds the most promise for constraining subsurface ice content on the Moon when used in conjunction with other datasets (such as thermal observations or nuclear spectroscopy) and when in the form of a gravimetry survey from a portable measurement device.

\subsection{Science objective \#4 -- Constrain the regional density of the crust}

\begin{figure}[!h]
\centering
\includegraphics[width=0.9\textwidth]{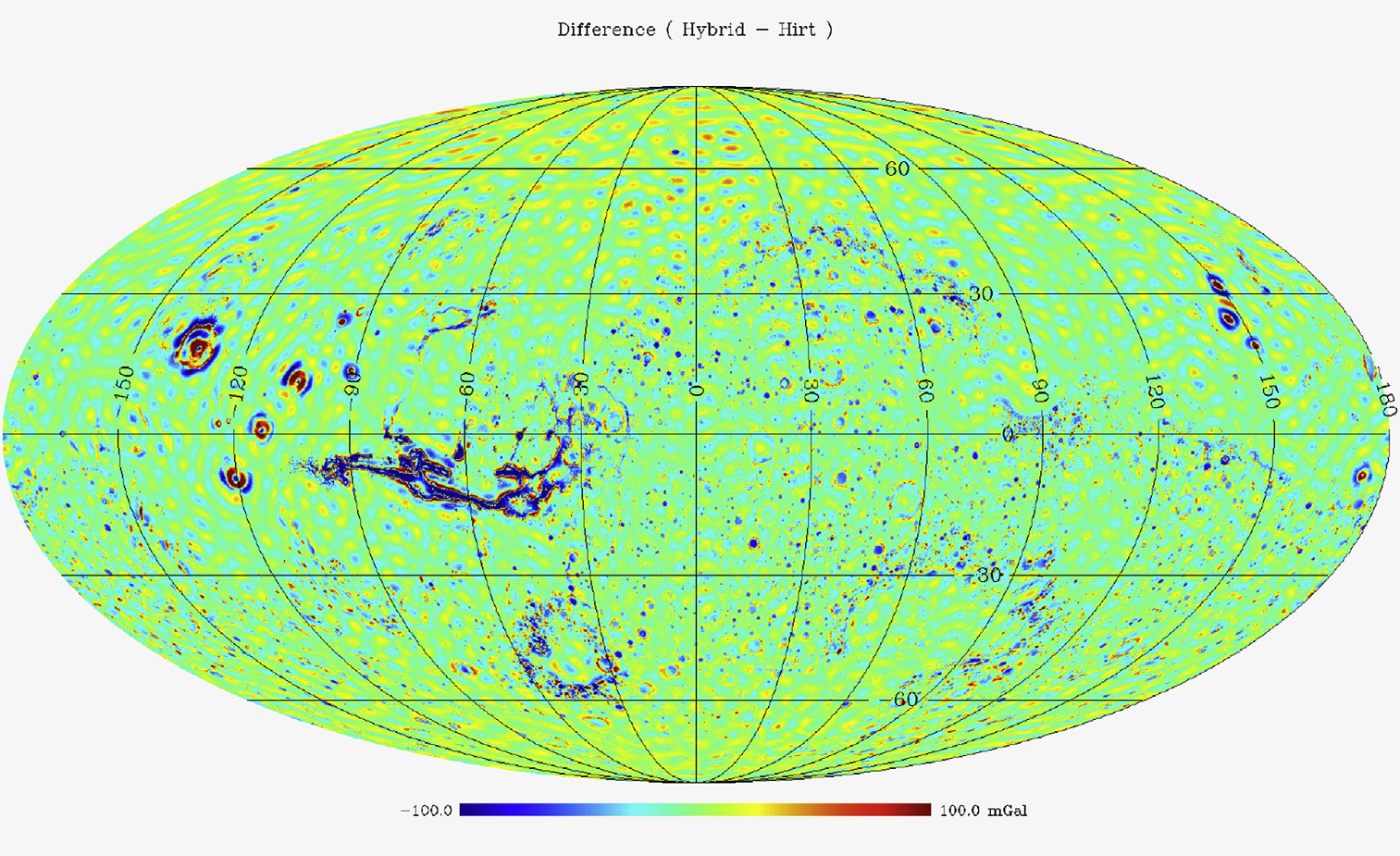}
\caption{From \citet{gorski2018high}, showing the difference in the calculated surface gravity on Mars resulting from different assumptions about crustal density. Note that the colorbar spans hundreds of mGals.}
\label{fig: Brillouin}
\end{figure}

When a planet's gravity field is sampled by an orbiting spacecraft, the gravitational potential at a given orbital altitude can be inferred by the potential at a different altitude through a process called continuation.  However, this kind of extrapolation is only possible outside the ``Brillouin sphere'' (defined to be the sphere that circumscribes all topography on the body).  Inside the Brillouin sphere, gravity is strongly dependent on the density of the crust.

This introduces an interesting opportunity for scientific investigation: if we were to measure absolute gravity on the surface of the Moon, this data point could be used to discern between different models of crustal density.  Different density distributions produce dramatically different surface gravity values, as shown in Figure \ref{fig: Brillouin} for Mars.  Slightly different assumptions about surface density can produce gravity accelerations that differ by hundreds of mGals.

\begin{figure}[!h]
\centering
\includegraphics[width=0.95\textwidth]{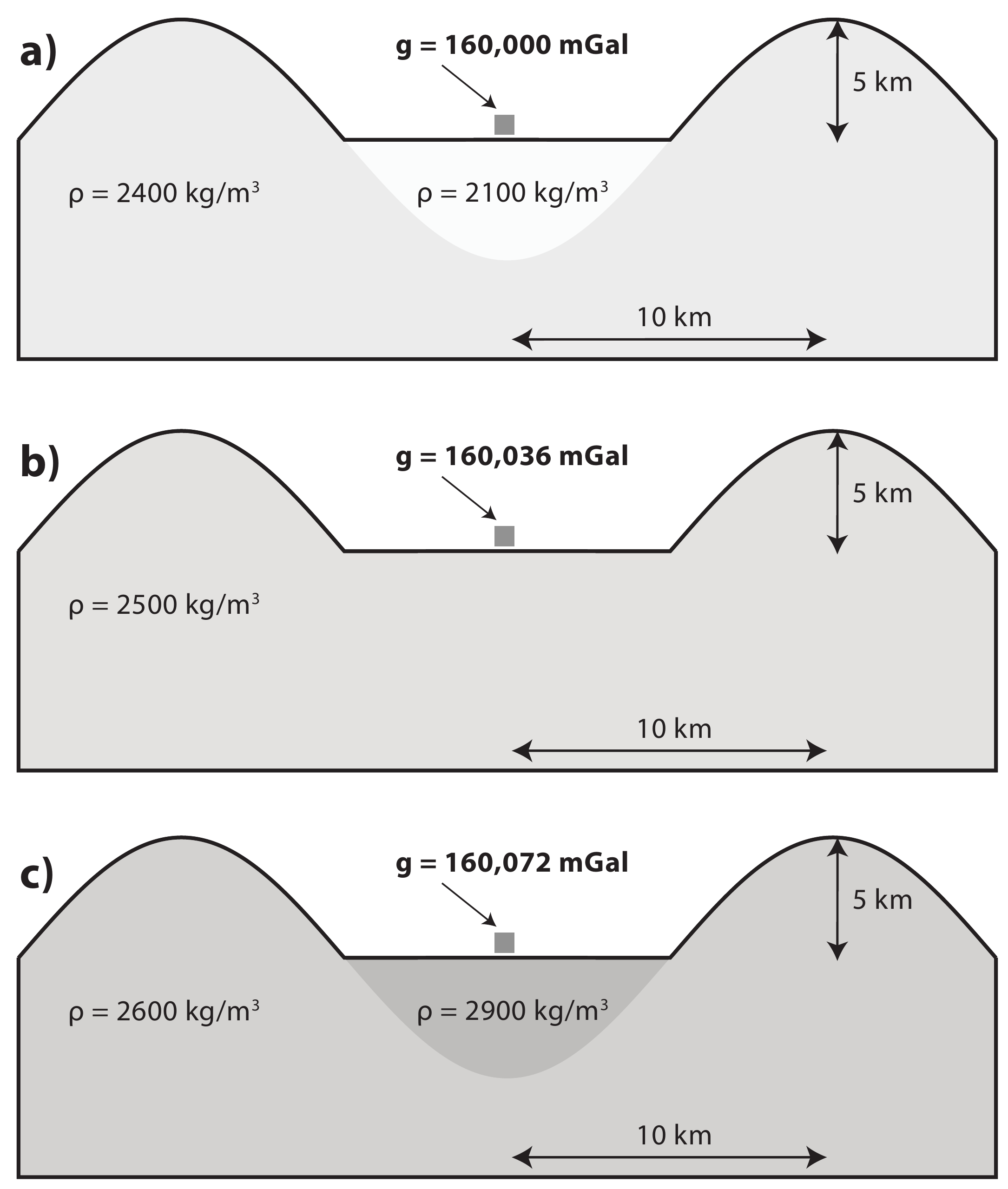}
\caption{Numerical calculations of surface gravity measurements for three density profiles. The gravity anomalies measured from orbit (at an altitude of 10 km) above each of these profiles are equivalent to within 1 mGal. Nevertheless, the gravity anomalies measured on the ground (in locations indicated by the small gray squares) vary by tens of milliGals.}
\label{fig:BrillouinGrav}
\end{figure}

We can approximate the impact of an absolute gravity measurement inside the Brillouin sphere with a 2-D numerical calculation.  Figure \ref{fig:BrillouinGrav} shows three density profiles that would produce a nearly identical gravity anomaly outside of the Brillouin sphere (i.e., at a constant altitude above the terrain).  Even though measurements of gravity from orbit would be unable to distinguish between these scenarios, gravity measured at the center of the valley differs by tens of milliGals. This is a consequence of what is known as a ``terrain effect'' in terrestrial gravimetry: nearby topography that looms above a gravimeter will contribute a pull away from the center of the planet, thereby decreasing the gravity reading. This is an effect that can not be completely predicted from measurements above the terrain.  This situation is analogous to the measurement of gravity inside the Moon's Brillouin sphere.

The density profiles in Figure \ref{fig:BrillouinGrav} would have vastly different implications for the geophysical context of the landing site.  Most models of short-wavelength GRAIL data assume a porosity profile akin to that of Figure \ref{fig:BrillouinGrav}b, i.e., uniform density.  However, different density profiles may be expected in reality.  Figure \ref{fig:BrillouinGrav}a illustrates a scenario in which low-lying topography has a relatively low bulk density, as is the case for many simple craters on the Moon.  Figure \ref{fig:BrillouinGrav}c shows the scenario in which high-density rocks infill low-lying terrains, as is generally the case for mare volcanism.

In the case of the density profiles shown in Figure \ref{fig:BrillouinGrav}, plausible measurements have a range in excess of 70 mGals.  The measurement requirement for this science objective must be tailored to a given landing site, but is reasonable to assume that an absolute gravity measurement at the lunar South Pole with an accuracy better than \textbf{20 mGals} would be a unique and valuable data point.

\subsection{Science objective \#5 --  Resolve deep internal structure}

\subsubsection{Tide-raising potential}

The solid body of the Moon flexes over time in response to gravitational interactions with Earth and the Sun.  These interactions produce a tide-raising (or ``inducing'') potential that would be felt by a gravimeter on the surface of the Moon. The main sources of tide-raising potential are: (1) the changing spacing of Earth and the Moon; (2) the relative position of the Sun; and (3) libration of the Moon. The first source produces the strongest variations in tide-raising potential, as quantified by \citet{qin2014perturbation} among others. The total range of accelerations that would be experienced at the Moons South Pole from the tide-raising potential would be just over 1 mGal. 

Although this is the range of gravity that would likely be detected by a gravimeter on the south pole of the Moon, this signal is easily predicted and not geologically interesting.  Rather, we want to design measurement requirements that will resolve the tidal response of the solid Moon, as described below.

\subsubsection{Solid body tidal response}

While the previous section describes the tide-raising potential that acts on the Moon and on gravimeters, the deformation of the solid Moon also results in gravity perturbations that can be measured by a gravimeter at the south pole.  These perturbations come from two effects: (1) rearrangement of mass within the Moon, which produces a new ``induced'' potential, and (2) radial displacement of the surface, which changes the distance of a gravimeter from the Moon's center of mass and thereby changes the strength of gravity acceleration.

The induced potential is quantified with the $k_{lm}$ Love numbers, where $l$ is spherical harmonic degree and $m$ is spherical harmonic order.  Love numbers are integral parameters of the internal structure and, unlike the static gravity field, are sensitive to the viscoelastic structure of the body.  For simple models of internal structure, the tidal response is solely dependent on the spherical harmonic degree; in other words, each term $k_{2m}$ may be approximated with a single degree-2 term, $k_2$. This effect corresponds to roughly 25 $\mu$Gals of gravity swings at the south pole.  For more complicated models of internal structure, the Moon can respond differently at different spherical harmonic orders. Note that a gravimeter at the Moon's south pole would specifically and exclusively detect the zonal ($m=0$) terms.

The solid surface of the Moon is displaced up and down as described by the $h_{lm}$ Love numbers.  Again, the tidal response would be purely a function of spherical harmonic degree for simple internal structure models. This tidal deformation would produce roughly 10 centimeters at the south pole over one tidal cycle, and this radial displacement corresponds to roughly 25 $\mu$Gals of gravity swings at the south pole. Note that this should approximately cancel out the gravity variations produced by the $k_{lm}$ Love numbers.  The total surface gravity variation is defined by the gravimetric parameter $\delta_{l} = 1 + 2h_{l}/l - (l+1)k_{l}/l$ and is given by:

\begin{equation}
\delta g^{tidal} = -2\frac{\delta_{2}}{R}U_{2}^{tidal}
\label{deltag}
\end{equation}

\noindent where $U_{tidal}$ is the tidal potential, and $R$ is the radius of the body. Eq. \ref{deltag} includes both inducing and induced tidal potential. However, only the induced potential depends on the internal structure.  In order to remove the inducing potential, we can modify the gravimetric factor as $\delta_{l}^{mod} = 2h_{l}/l - (l+1)k_{l}/l$. The measured values of the Love numbers are given in Table \ref{tab: LoveNumbersTable}.

\begin{table}[h!]
\begin{tabular}{|l|l|l|l|l}
\cline{1-4}
parameter & $k_{2}$                                                                                  & $h_{2}$                                                                                      & $\delta_{2}$ &  \\ \cline{1-4}
value     & \begin{tabular}[c]{@{}l@{}}0.02422$\pm$0.00022\\  \citep{williams2014lunar}\end{tabular} & \begin{tabular}[c]{@{}l@{}}0.0371$\pm$0.0033\\  \citep{mazarico2014detection}\end{tabular} & 1.00086$\pm$0.0033       &  \\ \cline{1-4}
\end{tabular}
\caption{Love numbers of the Moon considered in this section}
\label{tab: LoveNumbersTable}
\end{table}

\begin{figure}[!h]
\centering
\includegraphics[width=1.0\textwidth]{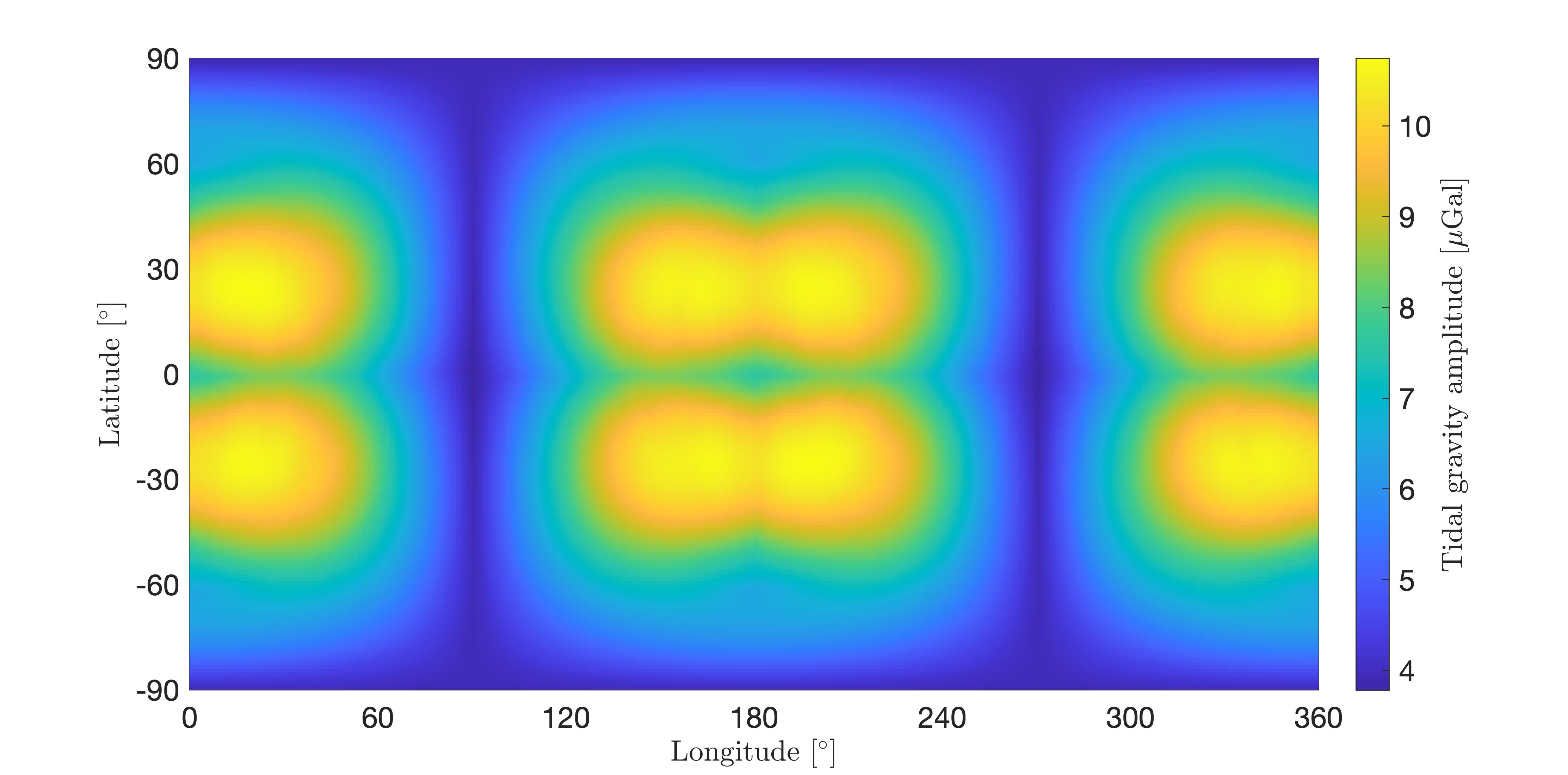}
\caption{Range of  tidally induced surface gravity variation associated with the Love numbers listed in Table \ref{tab: LoveNumbersTable}}
\label{fig: FigTidalRange}
\end{figure}

\begin{SCfigure}[\sidecaptionrelwidth][!h]
\includegraphics[width=0.4\textwidth]{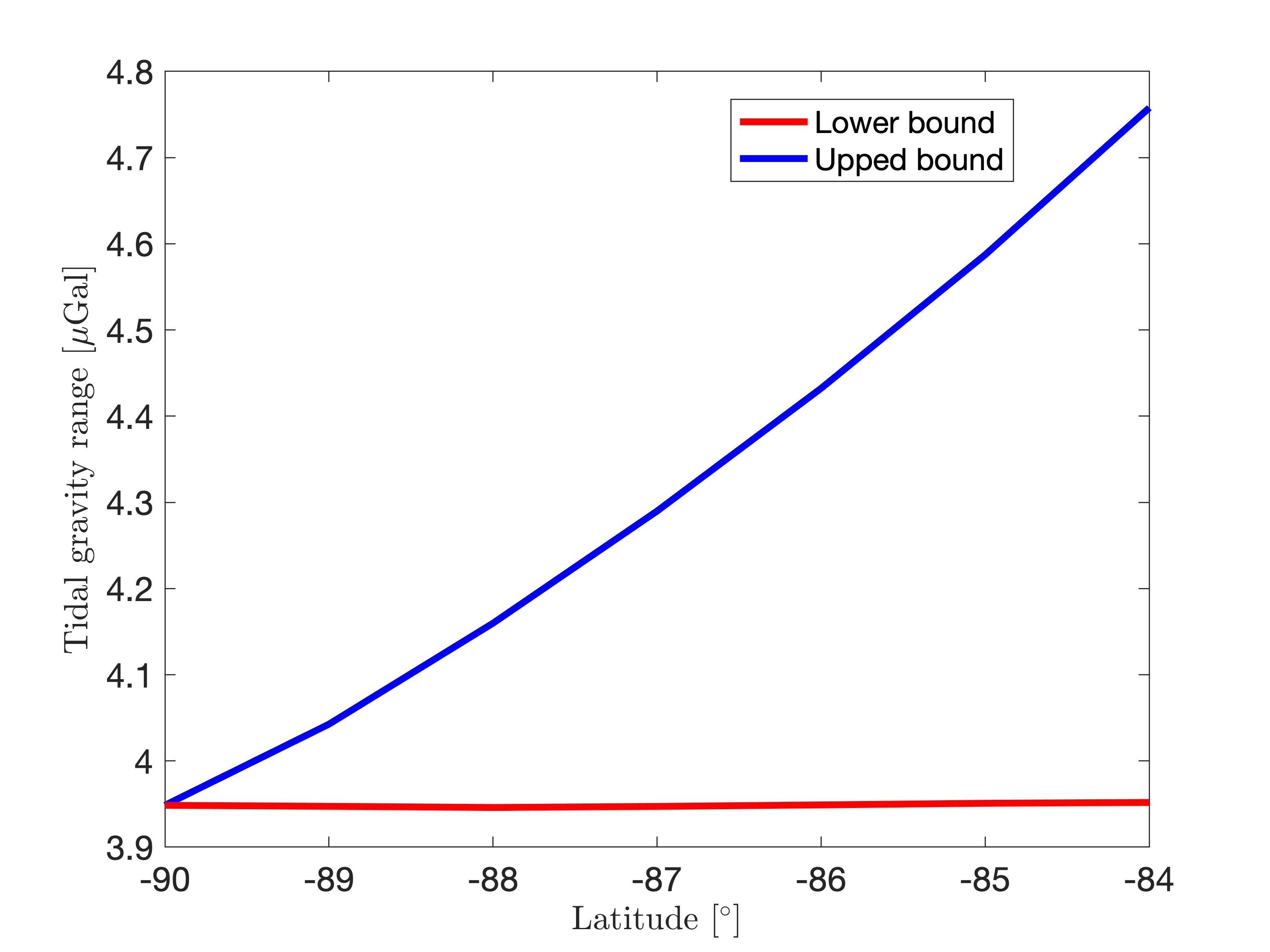}
\caption{The range of values that would be detected across all longitudes near the South Pole, for the model shown in Figure \ref{fig: FigTidalRange}.}
\label{fig:FigTidalRangeSouth}
\end{SCfigure}

We computed the induced degree-2 tidal surface gravity change due to the Moon and the Sun in the Moon Principal Axis (PA) coordinate system over the period of 18.6 years. Then, at each location, we found found the maximum and minimum value for the tidal surface gravity change. The difference between the maximum and the minimum value (i.e. tidal surface gravity range) is presented in Fig. \ref{fig: FigTidalRange} for the entire surface and in Fig. \ref{fig:FigTidalRangeSouth} for the vicinity of the South Pole. Only tidally induced surface gravity changed was computed.

\subsubsection{Precision with which gravimeter measurements must be made for the tidal response}

A gravimeter on the Moon would measure the tidal response, which is dependent on our uncertain knowledge of $k_{20}$, $h_{20}$, and all higher order terms that might arise from the deep structure of the Moon \citep{qin2014perturbation}. The tricky question here is what the uncertainty might be, given that we do not have a precise knowledge of the Moon's structure to begin with.  There is currently disagreement in the literature regarding the Moon's $h_2$ term.  For example, \citet{williams2014lunar} adopts a value of 0.0424 from lunar laser ranging (LLR), in contrast to the value of 0.0371 assumed in this section from Lunar Orbiter Laser Altimeter (LOLA) crossovers. The uncertainty in the LOLA-derived value of $h_2$ along with the discrepancy with LLR-derived values could change the observed gravity range by roughly 6 $\mu$Gals at the South Pole.  There additional uncertainty in higher-order Love numbers. Consequently, we will recommend a precision of \textbf{2 $\mu$Gals} for this science objective.

\subsection{Science objective \#6 --  Detect the solid inner core}

\begin{SCfigure}[\sidecaptionrelwidth][!h]
\includegraphics[width=0.5\textwidth]{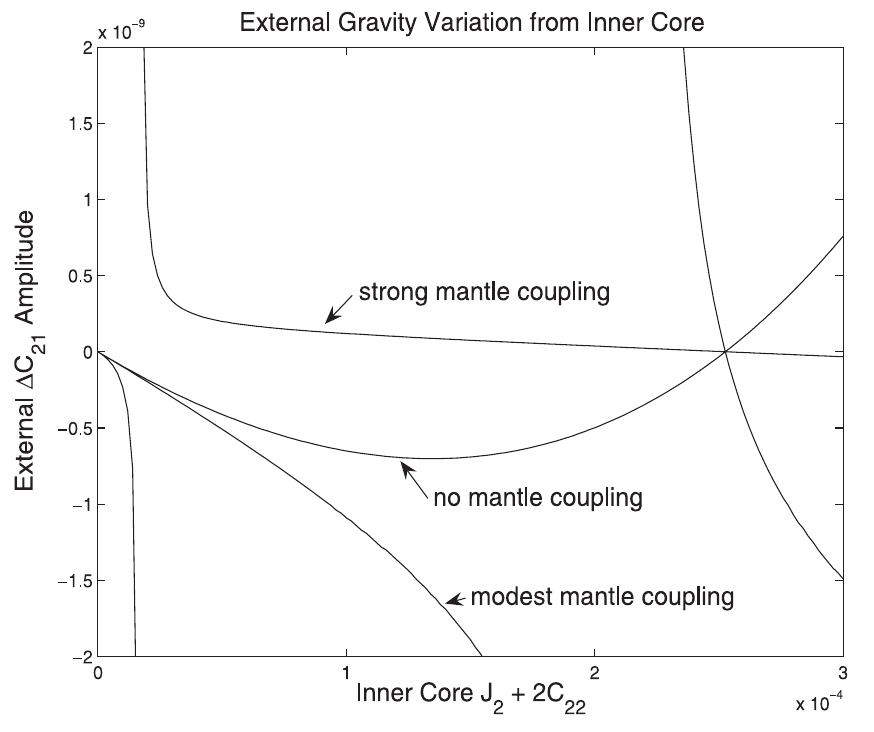}
\caption{Expected values for $\Delta C_{21}$ for various models, modified from \citet{williams2007scheme}}
\label{fig:Williams}
\end{SCfigure}

The Moon is expected to have a solid inner core, and the presence of a tilted inner core would be expected to produce a periodic variation in the degree 2 order 1 gravity coefficients \citep{williams2007scheme,dumberry2016forced}. As demonstrated by Figure \ref{fig:Williams} (from \citet{williams2007scheme}), some models for the inner core could be ruled out with a $C_{21}$ measurement precision of 5$\times 10^{-11}$.  A gravimeter at the South Pole would not be ideally situated to measure such variation, since the degree 2 order 1 harmonics are zero at the poles.  Nevertheless, an Artemis III landing site 6 degrees from the pole could allow detection of the inner core with a precision of 0.098 $\mu$Gals.  Consequently, a stationary gravimeter on Artemis III would need better than \textbf{0.1 $\mu$Gal} to plausibly address this science objective.  Note that this requirement would not need to be so stringent for landing sites at mid latitudes (20$^\circ$N--70$^\circ$N and 20$^\circ$S--70$^\circ$S).

\bibliographystyle{apalike}
\bibliography{main}
\end{document}